\documentclass[a4paper,11pt]{article}
\usepackage{pos}
\usepackage{bm}

\title{New Physics and Symmetry Tests with Polarized Photon Fusion and Dipole Moments}

\author*[a]{Fang Xu}

\affiliation[a]{Department of Physics, Center for Field Theory and Particle Physics, Fudan University,\\
  Shanghai, 200433, China}

\emailAdd{xufang@wustl.edu}

\abstract{We discuss new-physics searches and symmetry tests with dipole moments, emphasizing the role of polarization observables. As a primary benchmark, we consider polarized photon fusion in the $e^+ e^-$ environment of the Super Tau-Charm Facility (STCF) and study $\gamma \gamma \to \tau^+ \tau^-$ in the nearly back-to-back region, where a transverse-momentum-dependent (TMD) description provides a convenient framework for organizing polarization effects. We show that linearly polarized photons induce characteristic azimuthal asymmetries in the $\tau^+ \tau^-$ kinematics, enabling polarization-based observables that enhance sensitivity to the $\tau$ electromagnetic dipole form factors. Moreover, $CP$-even and $CP$-odd dipole interactions can be disentangled through distinct angular structures, offering a systematic path to probe $\tau$ dipole moments with improved precision at future lepton colliders. As an illustration, we obtain an improved $2\sigma$ reach on the anomalous magnetic dipole moment, $-4.6 \times 10^{-3} < \mathrm{Re}(a_\tau) < 7.0 \times 10^{-3}$, reaching a precision level close to the Standard Model expectation. To place these prospects in a broader context, we briefly summarize the experimental status of dipole-moment measurements across different fermionic systems and highlight their complementarity in constraining new physics. We illustrate this interplay with supersymmetric scenarios featuring $R$-parity violation, in which loop-induced dipole moments provide correlated probes of $CP$-conserving and $CP$-violating interactions. Taken together, polarized photon fusion and precision dipole measurements constitute a coherent program for testing fundamental symmetries and exploring physics beyond the Standard Model.}

\FullConference{
The 26th international symposium on spin physics (SPIN2025)\\
21–26 September, 2025\\
Qingdao (Tsingtao), Shandong Province, China\\}

\begin{document}
\maketitle

\section{Introduction}
Dipole moments provide a remarkably clean window into both precision tests of the Standard Model (SM) and searches for physics beyond the Standard Model (BSM). For a fermion, the anomalous magnetic dipole moment (MDM) is $CP$-even, whereas the electric dipole moment (EDM) is $CP$-odd, and the two therefore probe complementary aspects of the underlying dynamics. More broadly, dipole moments measured across different fermionic systems form a coherent symmetry-test program, in which improved experimental reach can translate into sensitivity to high new-physics scales. Polarization observables often unlock symmetry information that is inaccessible in unpolarized rates. This is true not only for hadronic systems, where spin-dependent distributions and correlations are standard tools, but also for purely leptonic processes once suitable polarization handles are available. In particular, the same logic behind azimuthal spin asymmetries~\cite{STAR:2019wlg,Li:2019yzy,Shao:2023bga} can be carried over and applied directly to electroweak reactions~\cite{Shao:2025xwp}, where polarization correlations provide clean access to $CP$-even and $CP$-odd structures. Among charged leptons, the $\tau$ lepton is particularly well motivated as a new-physics target, since the sensitivity of dipole moments to heavy new physics grows approximately with the lepton mass~\cite{Giudice:2012ms}. At the same time, $\tau$ dipole moments are experimentally challenging, because the short lifetime precludes the direct measurement strategies that work for the electron and muon; one must instead exploit production and decay observables in collider environments~\cite{Belle:2021ybo,CMS:2024qjo}.

In this work, we take polarized photon fusion to $\tau^+ \tau^-$ in an $e^+ e^-$ environment as a primary benchmark and use azimuthal-asymmetry observables to constrain the $\tau$ electromagnetic dipole form factors. We then place these prospects in a broader context by incorporating the latest experimental information on dipole-moment searches and by illustrating how concrete BSM scenarios can be tested through dipole observables. As a representative example, we briefly discuss supersymmetric models with trilinear $R$-parity violation (RPV), where loop-induced contributions to MDMs and EDMs yield correlated probes of $CP$-conserving and $CP$-violating interactions across different fermionic systems.

\section{Azimuthal asymmetries in polarized photon fusion in the correlation limit}

The photon-induced process $e^+(l_1)+e^-(l_2)\to e^+(l_1^\prime)+e^-(l_2^\prime)+\tau^+(p_1)+\tau^-(p_2)$ is mediated by the fusion of two quasi-real photons radiated from the incoming leptons, with momenta $k_1$ and $k_2$. We define the transverse momenta $\bm{P}_\perp\equiv(\bm{p}_{1\perp}-\bm{p}_{2\perp})/2$ and $\bm{q}_\perp\equiv\bm{p}_{1\perp}+\bm{p}_{2\perp}$. To analyze azimuthal observables in the correlation limit, we employ TMD factorization rather than the collinear framework commonly used in earlier two-photon studies. In close analogy to gluon TMDs in QCD~\cite{Pisano:2013cya}, the corresponding photon TMD PDFs in QED can be defined~\cite{Shao:2025xwp}
{\small
\begin{align}
    f_{\gamma}^{\alpha \beta}(x,\bm{k}_{\perp}) &= \int \frac{\mathrm{d} b^- \mathrm{d}^2 \bm{b}_\perp}{P^+ (2\pi)^3}e^{-i b^- (x P^+)+i \bm{b}_\perp\cdot \bm{k}_\perp} \left\langle e(P)\left|F^{+\alpha}(b) F^{+\beta}(0)\right| e(P)\right\rangle \big|_{b^+=0} \notag \\
    &= \ -\frac{g_{\perp}^{\alpha \beta}}{2} x f(x,\bm{k}_{\perp}^2) + \left(\frac{g_{\perp}^{\alpha \beta}}{2} + \frac{k_{\perp}^\alpha k_{\perp}^\beta}{\bm{k}_{\perp}^2} \right) x h_1^{\perp}(x,\bm{k}_{\perp}^2),
\end{align}
}
where $x$ is the photon longitudinal momentum fraction and $\bm{k}_\perp$ its transverse momentum. The transverse metric tensor is $g_{\perp}^{\alpha\beta}\equiv g^{\alpha\beta}-(n_a^\alpha n_b^\beta+n_a^\beta n_b^\alpha)$, with light-cone vectors $n_{a,b}^\mu=(1,0,0,\pm1)/\sqrt{2}$ and $b^\pm=(b^0\pm b^3)/\sqrt{2}$. The photon TMDs that describe unpolarized and linearly polarized photons are perturbatively calculable in QED~\cite{Bacchetta:2015qka,Shao:2025xwp}. The semi-inclusive di-tau differential cross section within the TMD factorization framework can thus be expressed as~\cite{Shao:2025xwp}
{\small
\begin{align}
& \frac{\mathrm{d} \sigma}{\mathrm{d}^2 \bm{P}_{\perp} \mathrm{d}^2 \bm{q}_{\perp} \mathrm{d} y_1 \mathrm{d} y_2} = \frac{1}{64 \pi^2 s^2} \int \mathrm{d}^2 \bm{k}_{1 \perp} \mathrm{d}^2 \bm{k}_{2 \perp} \delta^2\left(\bm{q}_{\perp}-\bm{k}_{1 \perp}-\bm{k}_{2 \perp}\right) x_1 x_2 \Big\{ \notag\\
&\big(\left|\mathcal{M}_{++}\right|^2+\left|\mathcal{M}_{--}\right|^2+\left|\mathcal{M}_{+-}\right|^2+\left|\mathcal{M}_{-+}\right|^2\big) f\big(x_1, \bm{k}_{1 \perp}^2\big) f\big(x_2, \bm{k}_{2 \perp}^2\big) \notag \\
&- 2\big[ \cos2(\phi_1-\phi_P) \mathrm{Re}\left( \mathcal{M}_{--} \mathcal{M}_{+-}^* + \mathcal{M}_{-+} \mathcal{M}_{++}^* \right) - \sin2(\phi_1-\phi_P) \mathrm{Im}\left( \mathcal{M}_{--} \mathcal{M}_{+-}^* + \mathcal{M}_{-+} \mathcal{M}_{++}^* \right) \big] \notag \\
&\quad \quad \times h_1^{\perp}\big(x_1, \bm{k}_{1 \perp}^2\big) f\big(x_2, \bm{k}_{2 \perp}^2\big) \notag \\
&- 2\big[ \cos2(\phi_2-\phi_P) \mathrm{Re}\left( \mathcal{M}_{++} \mathcal{M}_{+-}^* + \mathcal{M}_{-+} \mathcal{M}_{--}^* \right) - \sin2(\phi_2-\phi_P) \mathrm{Im}\left( \mathcal{M}_{++} \mathcal{M}_{+-}^* + \mathcal{M}_{-+} \mathcal{M}_{--}^* \right) \big] \notag \\
&\quad \quad \times f\big(x_1, \bm{k}_{1 \perp}^2\big) h_1^{\perp}\big(x_2, \bm{k}_{2 \perp}^2\big) \notag \\
&+ 2\big[ \cos2(\phi_1 - \phi_2) \mathrm{Re}\left( \mathcal{M}_{--} \mathcal{M}_{++}^* \right) - \sin2(\phi_1 - \phi_2) \mathrm{Im}\left( \mathcal{M}_{--} \mathcal{M}_{++}^* \right) \big] h_1^{\perp}\big(x_1, \bm{k}_{1 \perp}^2\big) h_1^{\perp}\big(x_2, \bm{k}_{2 \perp}^2\big) \notag \\
&+ 2\big[ \cos2(\phi_1 + \phi_2 - 2\phi_P) \mathrm{Re}\left( \mathcal{M}_{-+} \mathcal{M}_{+-}^* \right) - \sin2(\phi_1 + \phi_2 - 2\phi_P) \mathrm{Im}\left( \mathcal{M}_{-+} \mathcal{M}_{+-}^* \right) \big] \notag \\
&\quad \quad\times h_1^{\perp}\big(x_1, \bm{k}_{1 \perp}^2\big) h_1^{\perp}\big(x_2, \bm{k}_{2 \perp}^2\big)\Big\} ,
\label{eq:xsection2}
\end{align}
}
where $y_1$ and $y_2$ represent the rapidities of $\tau^+$ and $\tau^-$, $s \equiv (p_1 + p_2)^2$, and $\phi_P$ and $\phi_i$ denotes the azimuthal angle of $\bm{P}_{\perp}$ and $\bm{k}_{i \perp}$. $\mathcal{M}_{\lambda_1,\lambda_2}$ represents the helicity amplitude of the process $\gamma(k_1,\lambda_1) \gamma(k_2,\lambda_2) \to \tau^+(p_1) \tau^-(p_2)$ at $\phi_P=0$. We choose to retain and present this more cumbersome expression because, as will be shown later, it is highly relevant to the fundamental symmetry discussions that follow. Note that for parity-conserving cases where $\mathcal{M}_{\lambda_1,\lambda_2} = \mathcal{M}_{-\lambda_1,-\lambda_2}$, the terms $\mathcal{M}_{--} \mathcal{M}_{+-}^* + \mathcal{M}_{-+} \mathcal{M}_{++}^*$, $\mathcal{M}_{++} \mathcal{M}_{+-}^* + \mathcal{M}_{-+} \mathcal{M}_{--}^*$, $\mathcal{M}_{--} \mathcal{M}_{++}^*$ and $\mathcal{M}_{-+} \mathcal{M}_{+-}^*$ are real. This leads to the disappearance of the sine terms in Eq.~\eqref{eq:xsection2}. This actually implies that the condition of parity conservation causes the disappearance of potential $\sin{(2\phi)}$ and $\sin{(4\phi)}$ azimuthal asymmetries. Generally, such sine terms could arise from the imaginary part of such combinations that appeared in Eq.~\eqref{eq:xsection2}. As demonstrated below, this does contribute to the $\sin{(2\phi)}$ azimuthal asymmetry within the context of the study~\cite{Shao:2025xwp}. Retaining terms up to the second order in the dipole moments, the differential cross section can be expressed in the following form
{\small
\begin{align}
&\frac{\mathrm{d} \sigma}{\mathrm{d}^2 \bm{P}_{\perp} \mathrm{d}^2 \bm{q}_{\perp} \mathrm{d} y_1 \mathrm{d} y_2}\ =\ \frac{\alpha_e^2}{2 \pi^2 s^2}\Big[ \notag\\
&\quad \quad \big( C_0 + C_0^{\mathrm{Re}(F_2)} \mathrm{Re}(F_2) + C_0^{\mathrm{Re}(F_2)^2} \mathrm{Re}(F_2)^2 + C_0^{\mathrm{Re}(F_3)^2} \mathrm{Re}(F_3)^2 + C_0^{\mathrm{Im}(F_2)^2} \mathrm{Im}(F_2)^2 + C_0^{\mathrm{Im}(F_3)^2} \mathrm{Im}(F_3)^2 \big) \notag\\
&\quad +\big( C_{c2\phi} + C_{c2\phi}^{\mathrm{Re}(F_2)^2} \mathrm{Re}(F_2)^2 + C_{c2\phi}^{\mathrm{Re}(F_3)^2} \mathrm{Re}(F_3)^2 + C_{c2\phi}^{\mathrm{Im}(F_2)^2} \mathrm{Im}(F_2)^2 + C_{c2\phi}^{\mathrm{Im}(F_3)^2} \mathrm{Im}(F_3)^2 \big) \cos(2\phi) \notag\\
&\quad + C_{s2\phi}^{\mathrm{Im}(F_2)\mathrm{Im}(F_3)}\mathrm{Im}(F_2)\mathrm{Im}(F_3)\sin(2\phi) \notag\\
&\quad + \big( C_{c4\phi} + C_{c4\phi}^{\mathrm{Im}(F_2)^2} \mathrm{Im}(F_2)^2 + C_{c4\phi}^{\mathrm{Im}(F_3)^2} \mathrm{Im}(F_3)^2 \big) \cos(4\phi) \Big] ,
\label{eq:xsection3}
\end{align}
}
where $\phi \equiv \phi_q - \phi_P$. In this expansion, the form factors are treated as generally complex. The explicit expressions for the coefficients $C_{\mathcal{A}}^{\mathcal{B}}$ can be found in~\cite{Shao:2025xwp}. It is worth noting that the trigonometric functions in Eq.~\eqref{eq:xsection3} are inherited directly from the corresponding trigonometric terms in Eq.~\eqref{eq:xsection2} as a result of the Fourier expansion. As discussed earlier in this section, the cosine terms encode the parity-conserving contributions, whereas the sine terms encapsulate the parity-violating information. Based on the $C$, $P$, and $T$ transformation properties of the dipole form factors $F_{2,3}$, within the constant, $\cos(2\phi)$, and $\cos(4\phi)$ terms, the $CP$- and $T$-odd parameters, such as $\mathrm{Re}(F_3)$, $\mathrm{Re}(F_2)\mathrm{Re}(F_3)$, and $\mathrm{Im}(F_2)\mathrm{Im}(F_3)$ are prohibited; however, the case for $\sin(2\phi)$ is reversed. In fact, the terms $\mathrm{Re}(F_3)$, $\mathrm{Re}(F_2)\mathrm{Re}(F_3)$, and $\mathrm{Im}(F_2)\mathrm{Im}(F_3)$, although appearing in the intermediate steps of the constant or $\cos(2\phi)$ terms, are actually associated with coefficients $C_{\mathcal{A}}^{\mathcal{B}}$ that accompanied by sine functions rather than the cosine dependence. This sine dependence not only reflects their parity-odd nature but also ensures the vanishing of these coefficients upon completing the integration over the transverse momenta $\bm{k}_{1 \perp}$ and $\bm{k}_{2 \perp}$. The disappearance of these specific combinations in Eq.~\eqref{eq:xsection3} is not accidental but rather governed by the fundamental symmetry.

We propose three observables related to the $\cos(2\phi)$, $\sin(2\phi)$, and $\cos(4\phi)$ azimuthal angle asymmetry: $A_{c 2 \phi}=\frac{\sigma(\cos 2 \phi>0)-\sigma(\cos 2 \phi<0)}{\sigma(\cos 2 \phi>0)+\sigma(\cos 2 \phi<0)}$, $A_{s2\phi}=\frac{\sigma(y \sin 2 \phi>0)-\sigma(y \sin 2 \phi<0)}{\sigma(\sin 2 \phi>0)+\sigma(\sin 2 \phi<0)}$, and $A_{c 4\phi}=\frac{\sigma(\cos 4 \phi>0)-\sigma(\cos 4 \phi<0)}{\sigma(\cos 4 \phi>0)+\sigma(\cos 4 \phi<0)}$~\cite{Shao:2025xwp}. Notice that for the $A_{s2\phi}$ observable, the extra factor $y$ serves as a hemisphere weight since the $\sin(2\phi)$ modulation is odd under $y \to -y$. Notably, as discussed earlier, the $\sin(2\phi)$ term, and consequently, $A_{s2\phi}$, encapsulates the $CP$- and $T$-odd signatures of the cross section. The cancellation between the $y>0$ and $y<0$ regions exemplifies this symmetry, as the sign reversal of the total rapidity $y$ reflects the $CP$ and $T$ transformations of the system. Using the STCF~\cite{Achasov:2023gey} as a benchmark machine, the $\cos(2\phi)$ asymmetry yields constraints on $\mathrm{Re}(a_\tau)$ and $\mathrm{Re}(d_\tau)$ of the form
\begin{align}
    -4.6 \times 10^{-3}& < \mathrm{Re}(a_\tau) < 7.0 \times 10^{-3} ,\\
    |\mathrm{Re}(d_\tau)| &< 2.8 \times 10^{-16} \ e\,\mathrm{cm} ,
\end{align}
at $2\sigma$ CL. Notice that the $a_\tau$ reach demonstrates comparable sensitivity to the recent CMS measurement of $-4.2\times 10^{-3} < a_\tau < 6.2\times 10^{-3}$ at 95\% CL~\cite{CMS:2024qjo}, but without relying on the assumption about photon fluxes, and approaches the SM prediction of $a_\tau = 1.17721(5) \times 10^{-3}$. While the  $\sin(2\phi)$ and $\cos(4\phi)$ asymmetries lead to weaker bounds on $\tau$ dipole moments, they do not rely on the real parts of the corresponding form factors, and could therefore provide enhanced sensitivity to new-physics effects. A complete account of the analysis, including all kinematic selections, the implementation of $\tau$ decays, and detailed numerical results, is provided in~\cite{Shao:2025xwp}.

\section{Dipole-Moment Constraints on $R$-Parity-Violating Supersymmetry}
Trilinear RPV interactions in the Minimal Supersymmetric Standard Model (MSSM) provide a well-studied source of contributions to fermion dipole moments. In the standard setup adopted in the literature, where bilinear $R$-parity violation is neglected, and the usual supersymmetric CP phases, such as the $\mu$ parameter and the soft SUSY-breaking parameters, are set to zero, the RPV contributions to fermion dipole moments have been studied extensively; representative analyses can be found in~\cite{Kim:2001se,Altmannshofer:2025jkk}. A generic feature of this framework is that MDMs are generated already at one loop, while EDMs vanish at one loop and arise at two loops.

Because the sensitivity of dipole moments to heavy new physics grows approximately with the lepton mass~\cite{Giudice:2012ms}, this makes the $\mu$ and $\tau$ lepton especially well suited for dipole-based probes. In the muon sector, the $a_\mu$ has long served as a benchmark precision observable, and a large body of work has explored its implications for BSM physics and for specific frameworks such as RPV supersymmetry; see, e.g.,~\cite{BhupalDev:2021ipu,Afik:2022vpm} for representative studies. While recent experimental and SM updates have substantially reduced the apparent discrepancy $a_\mu^{\rm exp}-a_\mu^{\rm SM}=38(63)\times 10^{-11}$~\cite{Aliberti:2025beg}, corresponding to a $\sim 0.6\sigma$ difference, $a_\mu$ remains a benchmark precision observable for testing BSM scenarios, including RPV supersymmetry.

In~\cite{Altmannshofer:2025jkk}, it was found that, at the level of current experimental sensitivity~\cite{Muong-2:2008ebm,Belle:2021ybo}, the constraints derived from the $\mu$ and
$\tau$ EDM searches translate into $|\mathrm{Im}(\lambda^{(\prime)} \lambda^{*})| \lesssim \mathcal{O}(10^4)$ and $\mathcal{O}(10^6)$ for the $\mu$ and $\tau$, respectively, where $\lambda$ and $\lambda^\prime$ denote the trilinear $LLE$ and $LQD$ RPV couplings. This suggests that, with current sensitivities, the experimental limits on $d_\mu$ and $d_\tau$ still leave substantial room before they can impose truly competitive constraints on the RPV parameter space. Conversely, Table~\ref{tab:EDMconstraints} presents a target sensitivity estimate: fixing $m_{\rm SUSY}=1~{\rm TeV}$, we report the EDM reach in $d_\mu$ and $d_\tau$ that would be required for the corresponding bounds to push representative combinations of RPV couplings toward the perturbativity regime, i.e. to constrain $|\lambda^{(\prime)}\lambda^{*}|$ to be at $\sim \mathcal{O}(4\pi)$. Note that, since the CKM $CP$-violating phase is retained in our setup, the EDM constraints can also induce bounds on the real parts of certain RPV coupling combinations. It can be seen that, for the $d_\mu$ and $d_\tau$ measurements to impose meaningful constraints on the relevant RPV couplings, the current experimental sensitivities at the level of $10^{-19}$--$10^{-17}~e\,{\rm cm}$ would need to be improved to at least the $\mathcal{O}(10^{-23})~e\,{\rm cm}$ level (or below).

\begin{table}[htbp!]
\begin{tabular}{cc}
\hline\hline
$(|\mathrm{Im}(\lambda^{(\prime)} \lambda^*)|<4\pi,|\mathrm{Re}(\lambda^{(\prime)} \lambda^*)|<4\pi)$ & Corresponding $d_{e_k} (e\,\mathrm{cm}),(k=2,3)$ sensitivities \\ \hline
$\lambda_{i11} \lambda^*_{ikk},(i \neq 1,k)$                                                          & $(5.2 \times 10^{-26},\mathrm{N/A})$                     \\
$\lambda_{i22} \lambda^*_{ikk},(i \neq 2,k)$                                                          & $(6.6 \times 10^{-24},\mathrm{N/A})$                     \\
$\lambda_{i33} \lambda^*_{ikk},(i \neq 3,k)$                                                          & $(7.5 \times 10^{-23},\mathrm{N/A})$                     \\
$\lambda^\prime_{i11} \lambda^*_{ikk},(i \neq k)$                                                     & $(1.1 \times 10^{-25},\mathrm{N/A})$                     \\
$\lambda^\prime_{i22} \lambda^*_{ikk},(i \neq k)$                                                     & $(1.4 \times 10^{-24},\mathrm{N/A})$                     \\
$\lambda^\prime_{i33} \lambda^*_{ikk},(i \neq k)$                                                     & $(4.0 \times 10^{-23},\mathrm{N/A})$                     \\
$\lambda^\prime_{i12} \lambda^*_{ikk},(i \neq k)$                                                     & $(5.2 \times 10^{-28},1.7 \times 10^{-29})$              \\
$\lambda^\prime_{i13} \lambda^*_{ikk},(i \neq k)$                                                     & $(1.1 \times 10^{-25},5.1 \times 10^{-26})$              \\
$\lambda^\prime_{i21} \lambda^*_{ikk},(i \neq k)$                                                     & $(2.6 \times 10^{-29},8.4 \times 10^{-31})$              \\
$\lambda^\prime_{i23} \lambda^*_{ikk},(i \neq k)$                                                     & $(6.0 \times 10^{-25},1.2 \times 10^{-26})$              \\
$\lambda^\prime_{i31} \lambda^*_{ikk},(i \neq k)$                                                     & $(1.3 \times 10^{-28},5.7 \times 10^{-29})$              \\
$\lambda^\prime_{i32} \lambda^*_{ikk},(i \neq k)$                                                     & $(1.3 \times 10^{-26},2.6 \times 10^{-28})$              \\ \hline\hline
\end{tabular}
\caption{Required sensitivities in the $\mu$ and $\tau$ EDMs, to constrain representative combinations of trilinear RPV couplings toward the perturbativity regime, assuming $m_{\mathrm{SUSY}}=1~\mathrm{TeV}$.}
\label{tab:EDMconstraints}
\end{table}

\section{Conclusions}
We have highlighted how polarized photon fusion provides a clean handle for new-physics searches and symmetry tests through fermion dipole moments. Using the STCF as a benchmark, we showed that azimuthal-asymmetry observables in $\gamma\gamma\to\tau^+\tau^-$ within the TMD framework can improve the sensitivity to the $\tau$ dipole form factors, yielding a $2\sigma$ reach of $-4.6\times 10^{-3}<\mathrm{Re}(a_\tau)<7.0\times 10^{-3}$ and $|\mathrm{Re}(d_\tau)|<2.8\times 10^{-16}\,e\,\mathrm{cm}$. While the $\sin(2\phi)$ and $\cos(4\phi)$ asymmetries are typically less constraining numerically, their complementary dependence on the dipole form factors makes them particularly useful as diagnostics for nonstandard contributions. 

We also briefly discussed dipole-moment constraints on trilinear RPV supersymmetry. Current bounds on $d_\mu$ and $d_\tau$ are still far from probing perturbative values of representative RPV coupling combinations, indicating substantial room for future improvements. Overall, polarized photon fusion and precision dipole measurements constitute a coherent and complementary program to test fundamental symmetries and to explore physics beyond the Standard Model.

\bibliographystyle{JHEP}
\bibliography{ref}

\end{document}